\documentclass[prl,twocolumn,showpacs,preprintnumbers]{revtex4}
\usepackage{amssymb}

\usepackage{graphicx}
\usepackage{dcolumn}
\usepackage{bm}

\begin{document}

\title{Mott--Hubbard transition vs. Anderson localization of correlated,
disordered electrons}
\author{Krzysztof Byczuk,$^{1}$ Walter Hofstetter,$^2$ and Dieter Vollhardt$^3$ }
\affiliation{\centerline {$^1$ Institute of Theoretical Physics,
Warsaw University, ul. Ho\.za 69, PL-00-681 Warszawa, Poland, }
\centerline {$^2$ Condensed Matter Theory Group, Massachusetts Institute of Technology,
Cambridge, MA 02139, USA}\\
\centerline{$^3$ Theoretical Physics III, Center for Electronic Correlations and Magnetism,
Institute for Physics,}\\
\centerline{University of Augsburg, D-86135 Augsburg, Germany } }
\date{\today }

\begin{abstract}
The phase diagram of correlated, disordered electrons is
calculated within dynamical mean--field theory using the geometrically
averaged (''typical'') local density of states. Correlated metal, 
Mott insulator and Anderson insulator
phases, as well as coexistence and crossover regimes are
identified. The Mott and Anderson insulators are found to be
 continuously connected.
\end{abstract}

\maketitle




The properties of real materials are strongly influenced by the electronic
interaction and  randomness \cite{Lee85}.
 In particular, Coulomb correlations and disorder are both driving forces behind
metal--insulator transitions (MITs) connected with the localization and 
delocalization of particles. 
While the Mott--Hubbard MIT is caused
by the electronic repulsion \cite{Mott90}, the Anderson MIT is 
due to coherent backscattering  of non--interacting particles from randomly
distributed impurities \cite{Anderson58}. 
Furthermore, disorder and interaction effects are known to compete in subtle
ways \cite{Lee85,Kravchenko,ma}.
Several new aspects of this interplay will be discussed here. 

The theoretical investigation of disordered systems requires the use of
probability distribution functions (PDFs) for the random quantities of
interest. 
In physical or statistical problems one is usually interested in 
``typical'' values of random quantities which are  mathematically given by the most probable 
value of the PDF \cite{definition}.
In many cases the complete PDF is not known, i.e.,
only limited information about the system provided by certain averages 
(moments or cumulants) is available.
In this situation  it is of great importance to choose the most informative 
 average  of a random variable.
For example, if the PDF
of a random variable has a single peak and fast decaying tails this variable
is usually well estimated by its first moment, known as the \emph{arithmetic} average.
 However, there are many examples, 
e.g., from astronomy, the physics of glasses or networks, economy,
sociology, biology or geology, where the knowledge of the arithmetic average is insufficient 
since the PDF is so broad that its characterization requires infinitely many
moments. 
Such systems are said to be non--self--averaging. 
One example is Anderson localization: when a disordered system is
at the Anderson MIT \cite{Anderson58}, most of the electronic quantities fluctuate strongly
and the corresponding PDFs possess long tails \cite{Mirlin94}.
This is well illustrated by the local density of states (LDOS) of the
system. The arithmetic mean of this random one--particle quantity 
does not resemble its  typical value at all. 
In particular, it is non--critical at the Anderson transition \cite{Lloyd+Thouless} and hence
cannot help to detect the localization transition. By contrast, the 
\emph{geometric} mean \cite{lognormal,geometrical}, which gives a good
approximation of the most probable (``typical ``)
value of the LDOS, vanishes at a critical strength of the disorder and hence
provides  an explicit criterion for Anderson localization 
\cite{Anderson58,Dobrosavljevic97,Dobrosavljevic03,Schubert03}.

A non--perturbative theoretical framework for the investigation of correlated
lattice electrons with a local interaction is given  by dynamical
mean--field theory (DMFT) \cite{metzner89,georges96}.
If in this approach the effect of local disorder is taken into account
through the arithmetic mean of the LDOS \cite{ulmke95} one obtains, in the
absence of interactions, the well known coherent potential approximation
\cite{vlaming92}, which does not describe the physics of  Anderson
localization. To overcome this deficiency Dobrosavljevi\'{c} and Kotliar %
\cite{Dobrosavljevic97} formulated a variant of the DMFT where the
geometrically averaged LDOS is computed from the solutions of the
self--consistent stochastic DMFT equations. Employing a slave--boson
mean--field theory as impurity solver they investigated the disorder--driven
MIT for infinitely strong repulsion off half--filling. 
Subsequently, Dobrosavljevi\'{c} \emph{et al.} \cite{Dobrosavljevic03} 
incorporated the geometrically averaged LDOS into the self--consistency cycle and 
thereby derived  a mean--field theory of Anderson localization which
reproduces many of the expected features of the disorder--driven
MIT for non--interacting electrons. 
This scheme  uses only one--particle quantities and is therefore
easily incorporated into the DMFT for disordered electrons in the presence of 
phonons \cite{fehske}, or Coulomb correlations.

In this Letter we employ the DMFT with the typical LDOS to determine the
non--magnetic ground state phase diagram of the disordered Hubbard
model at half--filling for arbitrary interaction and disorder strengths. Thereby the
Mott--Hubbard and Anderson MITs are investigated on equal footing. The
system is described by a single--orbital Anderson--Hubbard model
\begin{equation}
H_{AH}=-t\sum_{\langle ij\rangle \sigma }a_{i\sigma }^{\dagger }a_{j\sigma
}+\sum_{i\sigma }\epsilon _{i}n_{i\sigma }+U\sum_{i}n_{i\uparrow
}n_{i\downarrow },  \label{1}
\end{equation}%
where $t>0$ is the amplitude for hopping between nearest neighbors, $U$ is
the on--site repulsion, $n_{i\sigma }=a_{i\sigma }^{\dagger }a_{i\sigma }^{{%
\phantom{\dagger}}}$ is the local electron number operator, $a_{i\sigma }$ 
($a_{i\sigma }^{\dagger}$) is the annihilation (creation) operator of an electron with spin $\sigma$,
and the local ionic energies $\epsilon _{i}$ are independent random variables. In the
following we assume a continuous probability distribution for $\epsilon _{i}$%
, i.e., $\mathcal{P}(\epsilon _{i})=\Theta (\Delta /2-|\epsilon
_{i}|)/\Delta ,$ with $\Theta $ as the step function. The
parameter $\Delta $ is a measure of the disorder strength. 

This model is solved within DMFT by mapping it \cite{georges96} onto an ensemble of
effective single--impurity Anderson Hamiltonians with different
$\epsilon _{i}$:
\begin{eqnarray}
H_{\mathrm{SIAM}} &=&\sum_{\sigma }(\epsilon _{i}-\mu )a_{i\sigma }^{\dagger
}a_{i\sigma }+Un_{i\uparrow }n_{i\downarrow }  \label{2} \\
&&+\sum_{\mathbf{k}\sigma }V_{\mathbf{k}}a_{i\sigma }^{\dagger }c_{\mathbf{k}%
\sigma }+V_{\mathbf{k}}^{\ast }c_{\mathbf{k}\sigma }^{\dagger }a_{i\sigma
}+\sum_{\mathbf{k}\sigma }\epsilon _{\mathbf{k}}c_{\mathbf{k}\sigma
}^{\dagger }c_{\mathbf{k}\sigma }.  \nonumber
\end{eqnarray}%
Here $\mu =-U/2$ is the chemical potential corresponding to a half-filled
band, and $V_{\mathbf{k}}$ and $\epsilon _{\mathbf{k}}$ are the
hybridization matrix element and the dispersion relation of the auxiliary
bath fermions $c_{\mathbf{k}\sigma }$, respectively. For each ionic energy 
$\epsilon _{i}$ we calculate the local Green function 
$G(\omega ,\epsilon _{i})$, from which we obtain the geometrically averaged LDOS 
$\rho _{\mathrm{geom}}(\omega )=\exp \left[ \langle \ln \rho _{i}(\omega )\rangle \right]$
\cite{Dobrosavljevic97,Dobrosavljevic03,comment2},
where $\rho _{i}(\omega )=-\mathrm{{Im}}G(\omega ,\epsilon _{i})/\pi $, 
and $\langle O_{i}\rangle =\int
d\epsilon _{i}\mathcal{P}(\epsilon _{i})O(\epsilon _{i})$ is the arithmetic
mean of $O_{i}$ \cite{remark1}. 
The lattice Green function is given by 
the corresponding Hilbert transform as $G(\omega )=\int d\omega ^{\prime }\rho _{\mathrm{%
geom}}(\omega )/(\omega -\omega ^{\prime })$.  
The local self--energy $\Sigma (\omega )$ is determined from the $\mathbf{k}$%
-integrated Dyson equation  
$\Sigma(\omega )=\omega -\eta (\omega )-G^{-1}(\omega )$ 
where the hybridization function $\eta (\omega )$ is defined as $%
\eta (\omega )=\sum_{\mathbf{k}}|V_{\mathbf{k}}|^{2}/\left( \omega -\epsilon
_{\mathbf{k}}\right) $. 
The self--consistent DMFT equations are closed through the Hilbert transform 
$
G(\omega )=\int d\epsilon N_{0}(\epsilon )/\left[\omega -\epsilon
-\Sigma (\omega )\right]
$, which relates the local Green function for a given lattice to the
self--energy; here $N_{0}(\epsilon )$ is the non--interacting DOS.
We note that this approach describes only the extended states
since the localized part of the spectrum, given by the isolated
poles of $G(\omega )$, is not included. For this reason $\rho_{\rm{geom}}(\omega)$ 
is not normalized to unity.

The Anderson--Hubbard model (\ref{1}) is solved for
a semi-elliptic DOS, $N_{0}(\epsilon )=2\sqrt{D^{2}-\epsilon ^{2}}/\pi D^2$,
with bandwidth $W=2D$; in the following we set $W=1$. For this DOS a simple
algebraic relation between the local Green function $G(\omega )$ and the
hybridization function $\eta (\omega )=D^{2}G(\omega )/4$ holds 
\cite{georges96}.
 The DMFT equations are solved
at zero temperature by the numerical renormalization group (NRG) technique %
\cite{NRG,bulla99} which allows us to calculate the geometric average of the LDOS in
each iteration loop.

\begin{figure}[tbp]
\includegraphics [clip,width=7.cm,angle=-00]{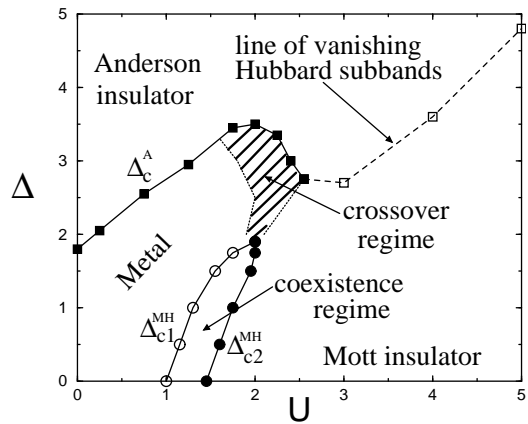}
\caption{Non--magnetic ground state phase diagram of the Anderson--Hubbard model at
half-filling as calculated by DMFT with the typical local density of states.
}
\label{fig1}
\end{figure}

\begin{figure}[tbp]
\includegraphics [clip,width=7.cm,angle=-00]{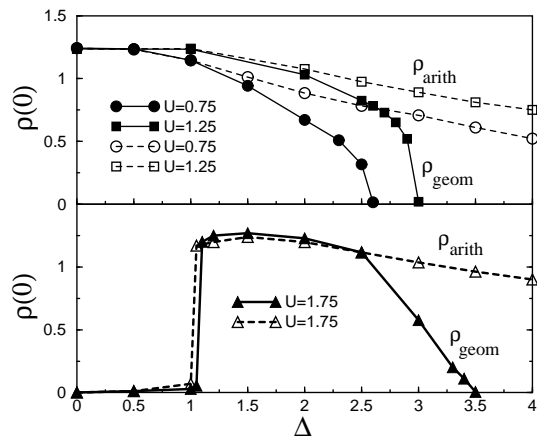}
\caption{Local density of states (LDOS) as a function of
disorder $\Delta$ for various values of the interaction $U$. Solid (dashed)
curves correspond to the geometrically (arithmetically) averaged LDOS.}
\label{fig2}
\end{figure}

\begin{figure}[tbp]
\includegraphics [clip,width=7.cm,angle=-00]{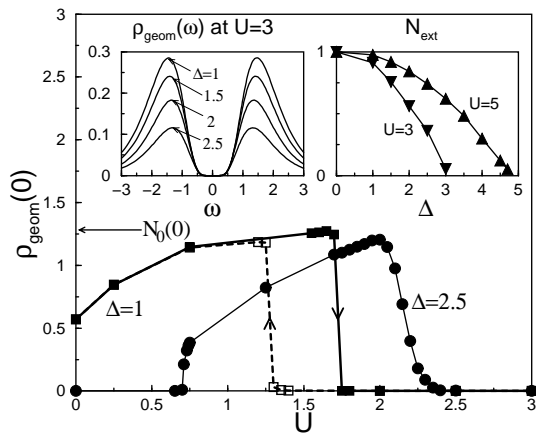}
\caption{Geometrically averaged LDOS as a function of
interaction $U$ for different disorder strengths $\Delta$. Solid (dashed)
curves with closed (open) symbols are obtained with an initial metallic
(insulating) hybridization function.  Triangles: 
$\Delta=1$; dots: $\Delta=2.5$.
Left inset: LDOS with Mott gap at $U=3$ for different disorder strengths $\Delta$.
Right inset: Integrated LDOS $N_{\rm{ext}}$ as a function of $\Delta$ at $U=3$
and $5 $.
}
\label{fig3}
\end{figure}

\begin{figure}[tbp]
\includegraphics [clip,width=7.cm,angle=-00]{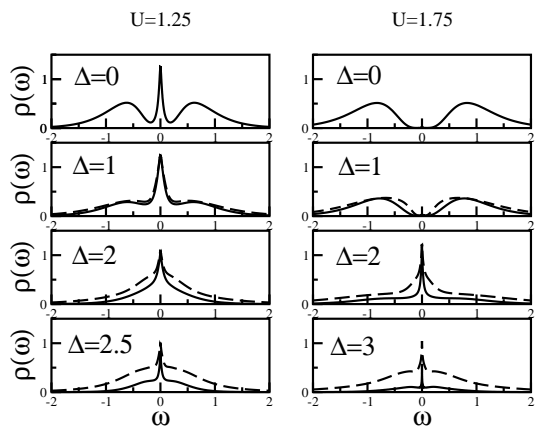}
\caption{LDOS for $U=1.25$ (left column)
and $U=1.75$ (right column) for different disorder strengths $\Delta$. Solid
(dashed) curves correspond to the geometrically (arithmetically) averaged LDOS.}
\label{fig4}
\end{figure}

The main result of this Letter is the ground state phase diagram of the
Anderson--Hubbard model at half-filling shown in Fig. \ref{fig1}. Two
different phase transitions are found to take place: a Mott-Hubbard MIT for
weak disorder $\Delta $, and an Anderson MIT for weak interaction $U$. The
two insulating phases surround the correlated, disordered metallic phase.
The properties of these phases, and the transitions between them, will now
be discussed.

\emph{(i) Metallic phase}: 
The correlated disordered metallic phase is characterized by a non--zero value of
 $\rho_{\rm{geom}}(0)$, the spectral density at the 
Fermi level ($\omega=0$). 
Without disorder  DMFT predicts  this quantity is to be given by the  
 bare DOS $N_0(0)$, as expressed by the  Luttinger theorem \cite{muller89}. 
This  means that Landau quasiparticles are well--defined at the 
Fermi level. 
The situation changes dramatically when randomness is introduced,
since a subtle competition between disorder and  electron interaction arises.
Increasing disorder at fixed $U$ reduces $\rho_{\rm{geom}}(0)$ and thereby decreases the metallicity
as shown in the upper panel of Fig. \ref{fig2}.
The opposite behavior is found   when  the interaction is
 increased at  fixed $\Delta$ (see Fig. \ref{fig3} for $\Delta=1$), i.e., the metallicity improves
in this case.
In the strongly interacting metallic regime
the value of $\rho_{\rm{geom}}(0)$  is restored, reaching again its maximal value $N_0(0)$. 
Physically this means that  in the metallic
 phase sufficiently  strong interactions protect the quasiparticles from their
 decay due to impurity scattering.
For weak disorder this effect  of the interaction is essentially  independent of the 
choice of the LDOS.

\emph{(ii)} \emph{Mott-Hubbard MIT}: For weak to intermediate disorder there
is a sharp transition at a critical value of $U$ between a correlated metal
and a gapped Mott insulator. We find two transition lines depending on whether
the MIT is  approached from the metallic side [$\Delta^{MH} _{c2}(U)$, full dots
in Fig. \ref{fig1}] or from the insulating side [$\Delta^{MH} _{c1}(U)$, open
dots in Fig. \ref{fig1}]. This is  very similar to the case without disorder
\cite{rozenberg94,georges96,bulla99};
the hysteresis is clearly seen  in Fig. \ref{fig3} for $\Delta =1$. 
The $\Delta^{MH}_{c1}(U)$ and $\Delta^{MH}_{c2}(U)$ curves in Fig. \ref{fig1} have positive slope.
This is a consequence of the disorder--induced increase  of spectral weight at the Fermi level (see
Fig. \ref{fig4}) which in turn requires  a stronger interaction to open the correlation gap. 
In the Mott insulating phase close to the hysteretic region 
an increase of disorder will therefore drive the system back into the metallic phase.
The corresponding abrupt rise of 
$\rho_{\rm{geom}}(0)$ is seen in the lower panel of Fig. \ref{fig2}
and the right column of Fig. \ref{fig4}.
In this case the disorder protects the metal from becoming  a Mott  insulator.

 Around $\Delta \approx 1.8$ the $\Delta^{MH}_{c1}(U)$
and $\Delta^{MH}_{c2}(U)$ curves terminate at a single critical point, cf. Fig. \ref{fig1}. 
At stronger disorder ($\Delta \gtrsim 1.8$) only a smooth crossover from a
metal to an insulator takes place. This is clearly illustrated by the $U$
dependence of $\rho_{\rm{geom}}(0)$ shown in Fig. \ref{fig3} for $\Delta =2.5$. In this
parameter regime the Luttinger theorem is not obeyed for any $U$. In the
crossover regime, marked by the hatched area in Fig. \ref{fig1}, $\rho_{\rm{geom}}(0)$
vanishes gradually, so that the metallic and insulating phases can no longer
be distinguished rigorously \cite{bulla01}. 

Qualitatively, we find again that 
the Mott-Hubbard MIT and the  crossover region do not depend much on the choice of the
average of the LDOS. We also note the similarity
between the Mott-Hubbard MIT scenario discussed here and the one for the
system without disorder at \emph{finite} temperatures \cite{georges93,rozenberg94,bulla01}, 
especially the presence of a coexistence region with hysteresis. However,
while in the non--disordered case the interaction needed to trigger the Mott-Hubbard
MIT decreases with increasing temperature, the opposite  holds in the
disordered case.

\emph{(iii) Anderson MIT}: In Fig. \ref{fig1} the metallic phase and the
crossover regime are seen to lie next to an
Anderson insulator phase where the LDOS  of the extended states
vanishes completely. 
The critical disorder $\Delta^A_{c}(U)$ corresponding to the  Anderson MIT  
 is a non--monotonous function of the interaction: it increases in the metallic regime and 
decreases  in the crossover regime.
Where $\Delta^A_c(U)$ has a positive slope an increase of the interaction  turns the 
Anderson insulator into a correlated metal. 
This is illustrated in Fig. \ref{fig3} for $\Delta=2.5$: at $U/W\approx 0.7$ a transition 
from a localized to a metallic phase occurs, i.e.,
 the spectral weight at the Fermi level becomes finite. 
In this case the electronic correlations impede the localization of quasiparticles due to 
impurity scattering.

 Fig. \ref{fig2} shows that the Anderson MIT is continuous. In the critical regime 
 $\rho_{\rm{geom}}(0)\sim \lbrack \Delta^A_{c}(U)-\Delta]^{\beta }$ for $U=\rm{const}$. 
In the crossover regime we find a critical exponent $\beta =1$ (see the
case $U=1.75$ in lower panel of Fig. \ref{fig2}); elsewhere $\beta \neq 1$. However, since
it is difficult to determine $\beta $ with high accuracy we cannot rule out
a very narrow critical regime with $\beta =1.$ It should be stressed that an
Anderson transition with vanishing $\rho_{\rm{geom}}(0)$ at finite $\Delta =\Delta^A_{c}(U)$
can only be detected in DMFT when the geometrically averaged LDOS is used (solid
lines in Fig. \ref{fig2}). With arithmetic averaging one finds a nonvanishing LDOS at
any finite $\Delta $ (dashed lines in Fig. \ref{fig2}).

\emph{(iv) Mott and Anderson insulators}: The Mott insulator with a
correlation gap is rigorously defined only for $\Delta =0$, and
the gapless Anderson insulator only for $U=0$ and $\Delta >\Delta^A_{c}(0)$.
In the presence of interaction and disorder this distinction can no
longer be made. However, as long as the LDOS shows the
characteristic Hubbard subbands (see left inset in Fig. \ref{fig3}) 
one may refer to a \emph{disordered Mott insulator}. With increasing 
$\Delta $ the spectral weight of the Hubbard subbands vanishes (see right
inset in Fig. \ref{fig3}) and the system becomes a \emph{correlated Anderson
insulator}. The border between these two types of  insulators is marked
by a dashed line in Fig. \ref{fig1}. The results obtained here within DMFT
prove that the paramagnetic Mott and Anderson insulators are  continuously connected. 
Hence, by changing $U$ and $\Delta$
it is possible to move from one type of insulator to the other without
crossing the metallic phase.

In conclusion, using DMFT with the geometrically averaged
(typical) LDOS we computed the non--magnetic ground state phase diagram of the
 Anderson--Hubbard model at half--filling
for arbitrary interaction and disorder strengths. 
In particular, we
determined the position of the Mott--Hubbard metal-insulator and
Anderson localization transitions. 
The presence of disorder  increases
the critical interaction  for the Mott-Hubbard MIT, and turns the
sharp transition (with hysteresis) into a smooth but rapid
crossover. On the other hand, the critical disorder strength  
for Anderson localization increases for weak interaction and is 
suppressed by strong interactions. The paramagnetic Mott and Anderson
insulators are continuously connected.
The specific predictions of our theory not only apply to disordered solids
but also to cold fermionic atoms in optical lattices \cite{optical}. 
In the latter case 
a precise control of system  parameters
appears to be possible which, in principle,  allows one to explore  
all parts of the phase diagram.

We thank R.~Bulla, S.~Kehrein, R.~Kutner, D.~Lobaskin, and J.~Tworzyd{\l}o
for useful discussions. This work was supported in part by the
Sonderforschungsbereich 484 of the Deutsche Forschungsgemeinschaft (DFG).
Financial support of KB through KBN-2 P03B 08 224, and of WH through the DFG
and a Pappalardo Fellowship is gratefully acknowledged.



\begin{references}
\bibitem{definition} The most probable value of a random quantity 
is defined as that value for which its PDF becomes maximal.
\bibitem{comment2} For a uniform PDF and $N$ discrete values of $\rho_i$ one finds 
$\rho_{\rm{geom}}=\left(\prod_{i=1}^N \rho_i\right)^{1/N}=
\exp[(\sum_{i=1}^N\ln \rho_i)/N]$. Hence $\rho_{\rm{geom}}$ 
vanishes if any of the $\rho_i$ is zero. The arithmetic average does not have such a property.
\bibitem{remark1} For numerical integrations  we use discrete values of $\epsilon_i$
selected according to the Gauss-Legendre algorithm \cite{ulmke95}.
\end{references}
\end{document}